\documentstyle[aps]{revtex}

\begin{document}

\title{Fingerprints of Classical Instability in Open Quantum Dynamics}

\author{
Paul A. Miller and Sarben Sarkar \\
{\it Department of Physics, King's College London, Strand, 
London WC2R 2LS, United Kingdom.}}

\maketitle

\bigskip

\begin{abstract}
The dynamics near a hyperbolic fixed point in phase space is modelled by an 
inverted harmonic oscillator. We investigate the effect of the classical 
instability on the open quantum dynamics of the oscillator, introduced 
through the interaction with a thermal bath, using both the 
survival probability function and the rate of von Neumann entropy increase, 
for large times. In this parameter range we prove, using influence functional
techniques, that the survival probability function decreases 
exponentially at a rate, $\kappa'$, depending not only on the 
measure of instability in the model but also on the strength of interaction 
with the environment. We also show that $\kappa'$ determines the rate of the 
von Neumann entropy increase and that this result is independent of the 
temperature of the environment. This generalises earlier results which are
valid in the limit of vanishing dissipation. The validity of inferring 
similar rates 
of survival probability decrease and entropy increase for quantum chaotic 
systems is also discussed.
\end{abstract}

\newcommand{\xii}{x_{i}}
\newcommand{\xip}{x'_{i}}
\newcommand{\xf}{x_{f}}
\newcommand{\xfp}{x'_{f}}
\newcommand{\hatom}{\hat{\omega}}
\newcommand{\hatgam}{\hat{\gamma_{0}}}
\newcommand{\PR}{{\em Physical Review }}
\newcommand{\PRL}{{\em Physical Review Letters }}

\section{Introduction}

Quantum-classical correspondence for the case of classically
chaotic systems has been much investigated recently (e.g., 
Refs.\cite{QC,Reichl,OdeA,Gutz,Haakebook,Houches}). 
Expectation values of 
corresponding dynamical variables begin to differ \cite{Casati,kickedtop}, 
as do classical and quantum phase space distributions \cite{Berry,Zurek1}, 
on extremely short timescales which are typically 
logarithmic in Planck's constant. If these studies are taken at face 
value, therefore, all chaotic dynamical systems, being fundamentally quantum 
in nature, should either be obeying quantum laws of evolution now or be 
expected to do so in an extremely short time. Observations tell us
otherwise, however. 

Sarkar and Satchell \cite{Sarben}, a decade ago, already pointed out the 
possible role of
environment in the quantum evolution of chaotic systems. Recently
\cite{Zurek2}, Zurek and Paz
have conjectured an interesting quantitative relation between a classical
chaotic system and its quantum version which is in contact with a bath. They 
have considered the Wigner representation of the quantum Liouville equation
\begin{equation}
\dot{W} = \left\{H,W\right\}_{PB} + \sum_{n=1}^{\infty} \frac{\hbar^{2n}
(-1)^{n}}{2^{2n} (2n+1)!} \frac{\partial^{2n+1}V}{\partial x^{2n+1}} 
\frac{\partial^{2n+1}W}{\partial p^{2n+1}}
\label{wigner1}
\end{equation}
for a particle in a potential $V(x)$ moving in a 2-dimensional phase space. 
Clearly the $\hbar$ terms are a singular perturbation of the classical
Liouville equation, in that the order of the differential equation is changed. 
For chaotic systems derivatives of the Wigner function 
with respect to momentum become large enough to render the quantum
correction terms comparable in magnitude to the classical Poisson bracket 
after the {\em Ehrenfest time}, $\tau_{\hbar} \propto 
(1 / \lambda) \log(1/\hbar)$, where $\lambda$ is a Lyapunov exponent. 
Therefore, for $t > \tau_{\hbar}$ we expect 
significant differences between the classical and quantum descriptions of the
same system. 

We will now consider 
an environment consisting of harmonic oscillators to be coupled to the
particle. The Caldeira-Leggett model \cite{Leggett,Book} will be used for these
oscillators. 
For the special case of a high-temperature, Ohmic environment the right hand
side of eqn.(\ref{wigner1}) is modified by the addition of a dissipitive and 
a decoherence 
term \cite{Zurek2}. When the temperature is 
high enough and the dissipation sufficiently low the dissipative term may be 
considered unimportant and the decoherence term only survives. 
The non-unitary Wigner function evolution becomes  
\begin{equation}
\dot{W} = \left\{H,W\right\}_{PB} + \sum_{n=1}^{\infty} \frac{\hbar^{2n}
(-1)^{n}}{2^{2n} (2n+1)!} \frac{\partial^{2n+1}V}{\partial x^{2n+1}} 
\frac{\partial^{2n+1}W}{\partial p^{2n+1}} + D \frac{\partial^{2}W}
{\partial p^{2}}.
\label{wigner2}
\end{equation}
The decoherence term is in the form of a diffusive contribution to the 
dynamics with diffusion coefficient, $D$. This is vital, since it is 
the diffusion resulting from the opening of
the system which limits the development of the fine structure in the
momentum direction to a critical momentum scale, $\sigma_{c}$. 
The timescale on which this process occurs is given by \cite{Zurek2}
\begin{equation}
\tau_{c} \approx \frac{1}{\lambda} \log \left( \frac{\sigma_{p}(0)}
{\sigma_{c}} \right),
\end{equation}
where $\sigma_{p}(0)$ is the initial width of a Gaussian wavepacket in the
momentum direction. Classical behaviour is recovered, therefore, provided 
the environmentally-induced diffusion process can prevent the development of
fine-structure, i.e. if 
\begin{equation}
\tau_{c} \ll \tau_{\hbar}.
\end{equation} 

However, opening a system to a thermal environment has other consequences as
well. In particular, the von Neumann entropy of the system, given by 
$S(t) = - \mbox{Tr} \rho_{r}(t) \ln \rho_{r}(t)$, where $\rho_{r}(t)$ is the 
reduced density matrix of the system at time $t$,  will increase;
information is lost to the environment and initially pure, superposition 
states of the quantum system become classical mixtures in a very short
time. 
Since our ability to predict accurately
the behaviour a classical system exposed to an initial perturbation 
depends very much on the nature of the dynamics, it is natural to ask whether
this is true for the rate of information loss of its quantum analogue due 
to a perturbing environment. 
 
As a first step towards answering this question (see also
\cite{CavesC,CavesQ}) Zurek and co-workers have
considered the inverted harmonic oscillator \cite{Barton} of unit mass with 
Hamiltonian
\begin{equation}
H_{S} = \frac{p^{2}}{2} - \frac{\lambda^{2} x^{2}}{2}.
\label{hamiltonian}
\end{equation}
This is intended as a model of instability and, in fact, the 
dynamical behaviour in phase space is dominated by a hyperbolic point at the 
origin. The unstable and stable directions, and the rate at which initial 
phase space distibutions expand and contract in these
directions, respectively, are determined by $\lambda$. In this sense we call 
$\lambda$ an ``instability
parameter" analagous to a Lyapunov exponent in a classical chaotic system. 
Indeed at any point on a trajectory the sum of the Lyapunov exponents is zero.
For a chaotic trajectory there must be one pair of non-zero Lyapunov exponents.

However, there are a number of reasons why we should question any conclusions
drawn as to the implications for a real chaotic system based on so simple a
model. Firstly, there are no quantum corrections to the Wigner function
evolution for this quadratic potential. The model does not allow for
these influences on the dynamics which, though small in the presence of an 
environment in comparison to the
classical terms, nonetheless are generally {\em always present}. 
The stable and unstable manifolds associated with all 
hyperbolic points in Hamiltonian chaotic systems intersect both one another 
and those associated with other hyperbolic points \cite{LL}. In this way 
homoclinic and
heteroclinic points are formed. The stable and unstable manifolds of the 
inverted oscillator intersect only at the hyperbolic origin in phase space. 
Clearly, therefore, the effect that the complicated distribution of homoclinic 
points might have on the open dynamics is not taken into account. Neither, 
of course, is the effect of heteroclinic points. 

Notwithstanding these objections however, the inverted oscillator remains a 
tractable model of instability both for a closed system and for an open system 
in the presence of an environment. As such, it deserves attention for the 
insights it might 
give regarding the qualitative and, maybe, quantitative behaviour of genuine, 
open quantum analogues of classically chaotic systems.      

The entropy production rate has been considered in the limit of high 
temperature and low dissipation. This entailed using the approximate Wigner
function evolution in eqn.(\ref{wigner2}). Zurek and Paz show 
\cite{Zurek2} that after a time
determined by both $\lambda$ and the strength of the interaction with the
environment the rate of entropy increase approaches a constant,
\begin{equation}
\dot{{\cal S}} \to \lambda,
\label{zurekrate}
\end{equation}    
i.e., the {\em quantum} entropy production rate is determined, in this
approximation, by the {\em classical} instability parameter. Given that the 
classical Lyapunov exponent to which $\lambda$ is analogous is equal to the 
Kolmogorov-Sinai (KS) entropy of the system \cite{Pesin}, this is indeed a 
remarkable
characterisation \cite{Miller}. It suggests that after a time, a quantum, 
classically 
chaotic system loses information
to the environment at a rate determined {\em entirely} by the rate at which the
classical system loses information as a result of its dynamics, namely the KS
entropy \cite{Beck}.       

In this paper we will examine once more the ``toy" model of Zurek and Paz. 
Apart from entropy production we will consider the experimentally relevant 
survival probability function. We will not be restricted to the 
assumption of low dissipation made by others \cite{Zurek2}. 
As we will show, the asymptotic behaviour of both the survival probability 
function and the rate of entropy increase will {\em not} be 
determined by $\lambda$ alone but by $\lambda$ in a specific combination with 
the dissipation parameter (which determines the strength of interaction with 
the environment). 
Our approach does not use the 
master equation approach of Zurek and Paz but rather Feynman-Vernon influence
functional techniques \cite{FV}. This allows a straightforward analysis of 
strong coupling of the system to the environment. 

The remainder of this paper is organised as follows:
In Section 2 we define the initial state of both the system and the 
environment, determine the time evolution of the reduced 
density matrix of the system generally and describe also how this allows us 
to calculate the von Neumann entropy, ${\cal S}(t)$. We specify the nature of 
the environment more explicitly in Section 3, leaving us in a position to 
consider finite temperature evolution. In Section 4 we shall define the 
survival probability function, 
$P(t)$, calculate it for the inverted oscillator and discuss its significance 
for quantum chaotic systems. 
In Section 5 we show analytically our generalisation of eqn.(\ref{zurekrate}) 
for the finite-temperature case. We state our conclusions in Section 6. 
Finally, the appendix contains the more tedious details of the calculations.

\section{Reduced Density Matrix Dynamics}

\subsection{Initial state}

We shall consider as our initial state the wavefunction
\begin{equation}
\psi(\xii,0) = (b \sqrt{\pi})^{-\frac{1}{2}} \exp \left(
\frac{-(\xii-x_{0})^{2}}
{2b^{2}} + ip_{0}\xii \right),
\label{wfninit} 
\end{equation}
for which it is easily verified that
\begin{eqnarray}
\langle \xii \rangle & = & x_{0}, \nonumber \\
\langle \xii^{2} \rangle & = & \frac{b^{2}}{2} + x_{0}^{2}, \nonumber \\
\langle p \rangle & = & p_{0}, \nonumber \\
\langle p^{2} \rangle & = & \frac{\hbar^{2}}{2b^{2}} + p_{0}^{2}, 
\quad\mbox{and} \nonumber \\
(\Delta \xii)^{2}(\Delta p)^{2} & = & \frac{\hbar^{2}}{4}, \nonumber 
\end{eqnarray}
i.e., a state of minimum uncertainty.
However, we shall be concerned with density matrices and their evolution so we
define the normalized initial density matrix corresponding to $\psi(\xii,0)$ 
above by
\begin{eqnarray}
\rho_{S}(\xii,\xip,0) & \equiv & \psi^{\ast}(\xii,0) \psi(\xip,0) \nonumber \\
& = & {\cal N} \exp ( -\epsilon ( \xii^{2} + {\xip}^{2} ) + 
a \xii + a^{\ast} \xip),
\label{initrho}
\end{eqnarray}
where the normalization constant $\cal{N}$ is given by 
\begin{equation}
{\cal N} = (b \sqrt{\pi})^{-1} \exp (-\frac{x_{0}^{2}}{b^{2}}),
\label{norm}
\end{equation}  
with $\epsilon = (2b^{2})^{-1}$ and $a = x_{0}/b^{2} + ip_{0}$.

We assume now that the system is put into contact with an environment
 at time $t = 0$. We will use as our environmental model a bath of independent
harmonic oscillators in thermal equilibrium at inverse temperature 
$1/(k_{B}T)$.
This allows us to write the initial, uncorrelated state
of the {\it combined} system in operator form as 
\begin{equation}
\hat{\rho}_{SE}(0) = \hat{\rho}_{S}(0) \otimes \hat{\rho}_{E}(0),
\label{initial}
\end{equation}
where 
\begin{equation}
\langle \xii | \hat{\rho}_{S}(0) | \xip \rangle = \rho_{S}(\xii,\xip,0)
\end{equation} 
as defined in eqn.(\ref{initrho}), and where $\hat{\rho}_{E}(0)$ defines the 
thermal environment,
\begin{equation}
\hat{\rho}_{E}(0) = \prod_{n} \left\{ \left[1 - \exp \left(- \frac{\hbar
\omega_{n}}{k_{B}T} \right) \right] \sum_{m} \exp \left(- \frac{m \hbar
\omega_{n}}{k_{B}T} \right) |m \rangle \langle m| \right\},
\end{equation}
i.e., in a factorized form because of the choice of non-interacting modes.

\subsection{Time evolution propagator}

The initial state being so defined, we now concentrate on the time evolution. 
We assume the Hamiltonian of the combined system to be
\begin{equation}
H_{SE} = H_{S} + H_{E} + H_{I},
\label{hamilt}
\end{equation}
where $H_{S}$ has been given in eqn.(\ref{hamiltonian}),
\begin{equation}
H_{E} = \sum_{n} \left( \frac{p_{n}^{2}}{2} + \frac{\omega_{n}^{2} 
q_{n}^{2}}{2} \right),
\end{equation}
i.e., the Hamiltonian of our chosen bath with canonical commutation relations 
$[q_{n},p_{m}] = i\hbar \delta_{n,m}$, and
\begin{equation}
H_{I} = - x \, c(t) \, \sum_{n} q_{n},
\end{equation}
the Hamiltonian of interaction describing the (possibly time dependent) 
coupling of the inverted oscillator, through its position variable, to the 
position variable of each of the environmental oscillators.

The combined system and environment, initially in the pure product state of 
eqn.(\ref{initial}), will, of course, evolve unitarily under the 
action of the Hamiltonian $H_{SE}$ of eqn.(\ref{hamilt}).   
We will be interested in the {\it reduced} density matrix of the system 
at some later time $t > 0$, which we write as $\hat{\rho}_{r}(t)$. 
To arrive at this expression the path integral method of
Feynman and Vernon is employed \cite{Leggett,FV}. The initially 
uncorrelated state allows us to 
calculate the evolution kernel for the reduced density matrix in the position 
basis in the following way:
\begin{equation}
\rho_{r}(\xf,\xfp,t) = \int d\xii \int d\xip \, J_{r}(\xf,\xfp,t|\xii,\xip,0) 
\rho_{S}(\xii,\xip,0),
\label{prop}
\end{equation}
where 
\begin{equation}
J_{r}(\xf,\xfp,t|\xii,\xip,0) = \int_{\xii}^{\xf} \mbox{D}x 
\int_{\xip}^{\xfp} \mbox{D}x' \, \exp 
\left\{\frac{i}{\hbar}(S[x]-S[x'])\right\} {\cal F}[x,x'],
\end{equation}
where $S$ is the action of the system and ${\cal F}$ is the influence
functional. Note that ${\cal F} = 1$ in the absence of an environment.
The propagator, $J_{r}$, has been calculated explicitly \cite{Hu1,Hu2} by 
Koks, Matacz and Hu and we quote here their result
\begin{eqnarray}
J_{r}(\xf,\xfp,t|\xii,\xip,0) &=& \frac{b_{2}}{2 \pi \hbar} \exp \left[
\frac{i}{\hbar}(b_{1} \Sigma_{f} \Delta_{f} - b_{2} \Sigma_{f} \Delta_{i}
+ b_{3} \Sigma_{i} \Delta_{f} - b_{4} \Sigma_{i} \Delta_{i}) \right. \nonumber
\\
& & \left. - \frac{1}{\hbar}(a_{11} \Delta_{i}^{2} + a_{12} \Delta_{f} 
\Delta_{i} + a_{22} \Delta_{f}^{2}) \right],
\label{kernel}
\end{eqnarray}
where we have used the more convenient sum and difference coordinates defined
by
\begin{equation}
\Delta \equiv x - x', \, \, \Sigma \equiv (x + x')/2,
\end{equation}
while $b_{1}, \ldots, b_{4}$ and $a_{11}, a_{12}$ and $a_{22}$ are time
dependent terms defined entirely by the spectral density and temperature of the
thermal environment. We will define them explicitly below when we specify the 
nature of the environment more precisely.

\subsection{Reduced density matrix evolution}

The final step in calculating the time evolution of the system is to use 
eqn.(\ref{kernel}) to calculate the reduced density matrix $\rho_{r}(t)$ 
in the position basis via eqn.(\ref{prop}) and eqn.(\ref{initrho}). 
One finds, after some lengthy but trivial algebra
\begin{eqnarray}
\rho_{r}(\xf,\xfp,t) &=& \frac{b_{2} {\cal N}}{\pi^{2} \sqrt{D}} \exp \left(
- \Gamma_{1} \Delta_{f}^{2} -\Gamma_{2} \Delta_{f} \Sigma_{f} - \Gamma_{3}
\Sigma_{f}^{2} \right. \nonumber \\ 
& & + \left. \Gamma_{5} \Sigma_{f}^{} + \Gamma_{6} \Delta_{f} + \Gamma_{4} 
\right),
\label{rhor}
\end{eqnarray}
where ${\cal N}$ has been defined in eqn.(\ref{norm}) above, 
\begin{equation}
D = 4 \hbar^{2} \epsilon^{2} + b_{4}^{2} + 8 \epsilon \hbar a_{11}, 
\label{dcoeff}
\end{equation}
and where we have made the abbreviations
\begin{equation}
\Gamma_{1} = \frac{a_{22}}{\hbar} + \frac{1}{D} \left\{
\left(\frac{\epsilon}{2} + \frac{a_{11}}{\hbar}\right)b_{3}^{2} +
\frac{b_{4} a_{12}b_{3}}{\hbar} - 2 \epsilon a_{12}^{2} \right\},
\label{Gam1}
\end{equation}
\begin{equation}
\Gamma_{2} = -2i \left\{\frac{b_{1}}{2\hbar} - \frac{1}{D} \left(\frac
{b_{2}b_{3}b_{4}}{2\hbar} - 2b_{2}a_{12}\epsilon \right) \right\},
\label{Gam2}
\end{equation}
\begin{equation}
\Gamma_{3} = \frac{2\epsilon b_{2}^{2}}{D}, 
\label{Gam3}
\end{equation}
\begin{equation}
\Gamma_{4} = {a^{\ast}}^{2}A_{7} + \frac{a^{2}\hbar^{2}}{D}\left(\epsilon -
A_{5}\right) + \frac{2 a_{11} \hbar a a^{\ast}}{D}, 
\end{equation}
\begin{equation}
\Gamma_{5} = x_{1}\left( \frac{2 a \hbar^{2}}{D}(\epsilon - A_{5}) + 
\frac{2 a_{11} \hbar a^{\ast}}{D} \right) + y_{1} \left(2 a^{\ast} A_{7} + 
\frac{2 a_{11} \hbar a}{D} \right),           
\end{equation}
and
\begin{equation}
\Gamma_{6} = x_{2} \left( \frac{2 a \hbar^{2}}{D}(\epsilon - A_{5}) + 
\frac{2 a_{11} \hbar a^{\ast}}{D} \right) + y_{2} \left(2 a^{\ast} A_{7} + 
\frac{2 a_{11} \hbar a}{D} \right),           
\end{equation}
but in which we have also defined
\begin{equation}
A_{5} = \frac{i b_{4}}{2 \hbar} - a_{11}{\hbar},
\end{equation}
\begin{equation}
A_{7} = \frac{1}{(\epsilon - A_{5})}\left(4 + \frac{a_{11}^{2}}{D} \right),
\end{equation}
\begin{equation}
y_{1} = x_{1}^{\ast} = \frac{i b_{2}}{\hbar},
\end{equation} 
\begin{equation}
x_{2} = \frac{i b_{3}}{2 \hbar} - \frac{a_{12}}{\hbar},
\end{equation}
and
\begin{equation}
y_{2} = \frac{i b_{3}}{2 \hbar} + \frac{a_{12}}{\hbar}.
\end{equation}

The expression in eqn.(\ref{rhor}) is of a Gaussian form and can be
diagonalized \cite{Zeh}. The von Neumann entropy, 
\begin{equation}
{\cal S}(t) = - \mbox{tr} \rho_{r}(t) \ln \rho_{r}(t),
\end{equation}
can therefore be calculated and written in the form 
\begin{equation}
{\cal S}(t) = - \frac{1}{p_{0}}\left(p_{0} \ln p_{0} + q \ln q \right),
\label{entropy}
\end{equation}
where $p_{0}$ and $q$ are defined by
\begin{equation}
p_{0} = \frac{2 \sqrt{\Gamma_{3}}}{\sqrt{\Gamma_{1}} + \sqrt{\Gamma_{3}}},
\label{p0}
\end{equation}
and
\begin{equation}
q = \frac{\sqrt{\Gamma_{1}} - \sqrt{\Gamma_{3}}}{\sqrt{\Gamma_{1}} + 
\sqrt{\Gamma_{3}}} = 1 - p_{0},
\label{q}
\end{equation}
using eqns.(\ref{Gam1}) and (\ref{Gam3}) above.

\section{Environment Specification}

\subsection{Generalities}

The influence functional used in the determination of the evolution kernel of
eqn.(\ref{prop}) is determined entirely by the so-called ``dissipation" and 
``noise" kernels of the chosen environment. If we now restrict each oscillator 
in the bath to have equal, unit mass we can write 
\begin{equation}
\mu(s,s') = - \int_{0}^{\infty} d \omega \, I(\omega,s,s') \sin(\omega(s-s'))
\label{mudef}
\end{equation}
and
\begin{equation}
\nu(s,s') = \int_{0}^{\infty} \, d \omega I(\omega,s,s') \coth \left(
\frac{\hbar \omega}{2 k_{B} T} \right) \cos(\omega(s-s'))
\label{nudef}
\end{equation}
for the dissipation and noise kernels, respectively, where 
\begin{equation}
{\cal F}[x,x'] = \exp \left\{ - \frac{1}{\hbar} \int_{0}^{t} \, ds 
\int_{0}^{s} \, 
ds' \Delta(s) \left[ \nu(s,s') \Delta(s') + 2 i \mu(s,s') \Sigma(s') \right]
\right\}.
\end{equation}
Here, $I(\omega,s,s')$ is called the {\it spectral density} of the environment 
as we have assumed the oscillators to have a continuous distribution of 
frequencies, $\omega$ \cite{Leggett}. Notice that $\mu(s,s')$ is independent of the
temperature of the environment. 

Restricting the discussion now to the case of the inverted oscillator we can
determine the time-dependent quantities $b_{1}(t), b_{2}(t), b_{3}(t)$ and 
$b_{4}(t)$, as well as $a_{11}(t), a_{12}(t)$ and $a_{22}(t)$, by
\begin{eqnarray}
b_{1}(t) &=& \dot{u_{2}}(t),  \nonumber \\
b_{2}(t) &=& \dot{u_{2}}(0),  \nonumber \\
b_{3}(t) &=& \dot{u_{1}}(t),  \nonumber \\
b_{4}(t) &=& \dot{u_{1}}(0) 
\label{bcoeffs}
\end{eqnarray}
and
\begin{equation}
a_{ij} = \frac{1}{1+\delta_{ij}} \int_{0}^{t} \, ds \int_{0}^{t} \, ds' 
v_{i}(s) \nu(s,s') v_{j}(s').
\label{acoeffs}
\end{equation}
The functions $u_{i}$ and $v_{i}$ which determine these quantities are 
solutions of the differential equations
\begin{eqnarray}
\ddot{u}(s) - \lambda^{2}u(s) + 2 \int_{0}^{s} \, ds' \mu(s,s') u(s') &=& 0, 
\label{udef} \\
\ddot{v}(s) - \lambda^{2}v(s) - 2 \int_{s}^{t} \, ds' \mu(s,s') v(s') &=& 0
\label{vdef}
\end{eqnarray}
when the boundary conditions
\begin{eqnarray}
u_{1}(0) = v_{1}(0) = 1, & & \, u_{1}(t) = v_{1}(t) = 0, \nonumber \\
u_{2}(0) = v_{2}(0) = 0, & & \, u_{2}(t) = v_{2}(t) = 1 \nonumber
\end{eqnarray}
are imposed.

\subsection{Calculation of $b_{i}(t)$}

It is clear from eqn.(\ref{bcoeffs}) and eqn.(\ref{udef}) above that the
quantities $b_{1}(t)$ to $b_{4}(t)$ depend only the dissipation kernel,
 $\mu(s,s')$, and not on the temperature of the environment.
We will concentrate on spectral densities of the form
\begin{equation} 
I(\omega,s,s') = \frac{2 \gamma_{0}}{\pi} \omega \exp \left( -
\frac{\omega}{\omega_{c}} \right) c(s) c(s'),
\label{ohmic}
\end{equation}
i.e., densities of an {\it Ohmic} type with an upper or cutoff frequency 
$\omega_{c}$ \cite{Leggett}. Using eqn.(\ref{ohmic}) in eqn.(\ref{mudef}) we find 
\begin{equation}
\mu(s,s') \to 2 \gamma_{0} c(s) c(s') \delta'(s-s')
\end{equation}
as $\omega_{c} \to \infty$.    
A further restriction to the case of constant coupling constants, i.e., $c(t) =
1 \, \forall t$, enables us to write for the inverted oscillator
\begin{eqnarray}
b_{1}(t) &=& (- \gamma_{0} + \kappa \coth\kappa t), \nonumber \\
b_{4}(t) &=& (- \gamma_{0} - \kappa \coth\kappa t), \nonumber \\
b_{2}(t) &=& \frac{\kappa \exp (\gamma_{0} t)}{\sinh \kappa t} \,\,
\,\,\,\, \mbox{and} 
\nonumber \\
b_{3}(t) &=& \frac{- \kappa \exp (- \gamma_{0} t)}{\sinh \kappa t}.
\label{bs}
\end{eqnarray}
The shorthand definition
\begin{equation}
\kappa = \sqrt{\lambda^{2} + \gamma_{0}^{2}}
\label{kappa}
\end{equation}
has been used here and we will see that it is an important quantity 
in the sections below. The asymptotic behaviour of $b_{i}$, $i = 1,\ldots,4$, 
in eqn.(\ref{bs}) for large $\kappa t$ can easily be determined and each is 
given by one of
\begin{eqnarray}
b_{1}(t) &\to& (- \gamma_{0} + \kappa), \label{b1asym} \\ 
b_{2}(t) &\to& 2 \kappa \exp (- (\kappa - \gamma_{0})t), 
\label{b2asym} \\
b_{3}(t) &\to& 2 \kappa \exp (- (\kappa + \gamma_{0})t),
\label{b3asym} \, \, \, \, \, \, \mbox{or} \\
b_{4}(t) &\to& (- \gamma_{0} - \kappa). \label{b4asym} 
\end{eqnarray}

Finally, $v_{1}(t)$ and $v_{2}(t)$ are required for the calculation of each 
$a_{ij}$ in eqn.(\ref{acoeffs}). Again for the inverted harmonic oscillator
they are given by
\begin{equation}
v_{1}(s) = \frac{\sinh(\kappa(t - s)) \exp(\gamma_{0}s)}{\sinh(\kappa t)}
\label{v1}
\end{equation}
and
\begin{equation}
v_{2}(s) = \frac{\sinh(\kappa(s)) \exp(\gamma_{0}(s - t))}{\sinh(\kappa t)},
\label{v2}
\end{equation}
for $s \in [0,t]$, where $t$ is the time at which we wish to calculate the 
reduced density matrix.

\subsection{Calculation of the finite temperature $a_{i}(t)$ coefficients}

To examine the asymptotic rate of entropy production of the inverted 
oscillator, in an environment at a {\em finite} 
temperature, we must, of course, calculate the approprate $a_{ij}$ 
coefficients. This will allow to analytically derive the asymptotic rate of 
von Neumann entropy increase.
The lengthy details are relegated to Appendix A. Writing 
$\hat{\lambda} := \lambda / \kappa$ we find eventually
\begin{eqnarray}
a_{11}(z,0) &=& \frac{\hatgam}{(2 \hat{\lambda} \sinh z)^{2}}
\sum_{n=0}^{\infty} f(n) \left(\frac{d_{1}(n) f_{1}(z)}{\hatgam} + 
d_{2}(n) f_{2}(z) \right), \label{finta11} \\     
a_{12}(z,0) &=& \frac{e^{-\hatgam z}}{2 (\hat{\lambda} \sinh z)^{2}}
\sum_{n=0}^{\infty} f(n) \left(d_{1}(n) f_{3}(z)- d_{2}(n) f_{4}(z) \right), 
\label{finta12} \\   
a_{22}(z,0) &=& \frac{e^{-\hatgam z}}{(2 \hat{\lambda} \sinh z)^{2}}
\sum_{n=0}^{\infty} f(n) \left(d_{1}(n) f_{5}(z) + 
d_{2}(n) f_{6}(z) \right), \label{finta22}
\end{eqnarray}
where the functions $f_{i}(z), i = 1,\ldots,6$ are defined in Appendix A, 
as are $d_{1}(n)$ and $d_{2}(n)$. Also given in Appendix A is the asymptotic
behaviour of each (finite-temperature) $a_{ij}$.

\section{Survival probability function}

\subsection{Definition and context}

In this section we will consider the {\em survival probability function}, 
$P(t)$, defined by
\begin{equation}
P(t) = \mbox{Tr}[\rho(0) \rho(t)],
\label{spf}
\end{equation}
for a system initially described by the density matrix $\rho(0)$. In
particular, we will be interested in the asymptotic behaviour.      

The function $P(t)$ has been examined before in the context of quantum chaos.
Tameshtit and Sipe \cite{Sipe} have considered its behaviour
in time for both regular and chaotic systems coupled in a non-demolition 
fashion to a high-temperature thermal reservoir. Using a master equation 
approach and random matrix theory they show that when suitably averaged, 
the behaviour of $P(t)$ displays substantial differences depending on the 
nature of the underlying dynamics. This function has also been examined 
theoretically for pure states \cite{Pechukas,Wilkie}, even for the inverted 
oscillator model \cite{Barton}. However, the fact that it is accessible 
experimentally \cite{spfexp} makes it a particularly important quantity to 
examine.

For {\em pure} initial states, $\rho(0) = |\phi(0) \rangle \langle \phi(0) |$, 
and unitary evolution, $|\phi(t) \rangle = \hat{U}(t) |\phi(0) \rangle$, we 
have 
\begin{equation}
P(t) = |\langle \phi(0) |\phi(t) \rangle|^{2},
\end{equation}
i.e., the probability of the system being in the initial state at a later 
time $t$. $P(t)$, as defined in eqn.(\ref{spf}) above, is a
generalisation of this applicable to systems which may not be
initially pure and/or for which the evolution in time is non-unitary.

\subsection{Survival probability for the open inverted oscillator}

We will now examine the effect of the thermal bath on the survival probability
function of the inverted harmonic oscillator. For simplicity we will 
consider our initial state to be centered at the origin of phase space, i.e.,
we take our initial wavefunction to be given by eqn.(\ref{wfninit}) with 
$x_{0} = p_{0} = 0$. The initial density matrix then given by 
eqn.(\ref{initrho}) with $a = 0$ and ${\cal N} = (b \sqrt{\pi})^{-1}$. These 
initial 
conditions simplify the form of the reduced density matrix considerably, 
leading to eqn.(\ref{rhor}) with $\Gamma_{4} = \Gamma_{5} = \Gamma_{6} = 0$.
The survival probablity function can now be simply calculated according to 
eqn.(\ref{spf}) above. We find 
\begin{eqnarray}
P(t) &=& \mbox{Tr}[\rho_{S}(0) \rho_{r}(t)] \nonumber \\
&=& \int_{-\infty}^{\infty} \, dx \int_{-\infty}^{\infty} \, dx' 
\langle x | \rho_{S}(0) | x' \rangle \langle x' | \rho_{r}(t) | x \rangle 
\nonumber \\
\ldots \, \, &=& \frac{1}{b \pi^{2}} \left\{\frac{\Gamma_{3}}{(\epsilon + 
2 \Gamma_{1})
(\epsilon + \Gamma_{3} / 2) - \Gamma_{2}^{2} / 4} \right\}^{\frac{1}{2}},
\label{spft}
\end{eqnarray}
where we have yet to specify the magnitude of the parameters of the 
environment. For the finite 
temperature case, we can easily determine the asymptotic behaviour. Using 
the asymptotic behaviour of each $b_{i}(t)$ (given by 
eqns.(\ref{b1asym}) - (\ref{b4asym})), the asymptotic behaviour of each 
$a_{ij}(t,0)$ (given in eqns.(\ref{ta11asym}), (\ref{ta12asym}) and 
(\ref{ta22asym})), and the 
definitions of each $\Gamma_{i}$ (eqns.(\ref{Gam1}), (\ref{Gam2}) 
and (\ref{Gam3})) we easily see that, for a fixed temperature
\begin{equation}
\left\{ (\epsilon + 2 \Gamma_{1})
(\epsilon + \Gamma_{3} / 2) - \Gamma_{2}^{2} / 4 \right\}^{\frac{1}{2}} 
\to C_{1},
\end{equation}
for large times, where $C_{1}$ is a (temperature-dependent) {\em constant}. 
Consequently, 
the large time behaviour of $P(t)$ is determined {\em entirely} by that of 
$\Gamma_{3}$. As $t(z) \to \infty$ we see from eqns.(\ref{b4asym}), 
(\ref{ta11asym}) and 
(\ref{dcoeff}) that $D$ goes to the (temperature-dependent) {\em constant}
\begin{equation} 
D_{asym} = 4 \hbar^{2} \epsilon^{2} + (\gamma_{0} + \kappa)^{2} - 
(4 \epsilon \hbar \gamma_{0}) \coth \left( \frac{(\gamma_{0} - \kappa) \hbar}
{2 k_{B} T} \right).
\label{dasym}
\end{equation}
Now, using its definition in
eqn.(\ref{Gam3}), the asymptotic behaviour of $b_{2}$ in eqn.(\ref{b2asym}) 
and that of $D$ in eqn.(\ref{dasym}) we find, finally,
\begin{equation}
P(t) \stackrel{t \to \infty}{=} C_{2} \exp\left(-(\kappa - \gamma_{0})t
\right),
\label{spfasym}
\end{equation}
where $C_{2}$ is another temperature-dependent constant. We have checked the 
accuracy of this result numerically by calculating $\log[P(t-1)/P(t)]$, which,
 if eqn.(\ref{spfasym}) is to be believed, has $\kappa - \gamma_{0} 
\equiv \kappa' = 
\sqrt{\lambda^{2} + \gamma_{0}^{2}} - \gamma_{0}$ as its asymptotic value. 
Excellent agreement was found. This is a generalisation of the result of 
Heller \cite{Houches} to the {\em open} quantum inverted oscillator and reduces 
to it as $\gamma_{0} \to 0$, i.e. as the system interacts more weakly with the
environment, as required. It would be natural to conjecture in the spirit of
Zurek and Paz \cite{Zurek2} that the behaviour in eqn.(\ref{spfasym}) might be
expected to hold for quantisations of classically chaotic systems in 
interaction with an environment.

\section{Asymptotic rate of entropy increase - analytical results}

We are now in a position to derive the rate at which the von 
Neumann entropy will increase when $z = \kappa t$ is large, i.e., at long times 
and/or when the Lyapunov exponent is large, for the finite temperature case. 

We can rewrite eqn.(\ref{p0}) as
\begin{equation}
p_{0} = \frac{2 \alpha}{1 + \alpha},
\label{p01}
\end{equation}
where we have made the abbreviation 
$\alpha := \sqrt{\Gamma_{3} / \Gamma_{1}}$. Also, eqn.(\ref{entropy}) gives
\begin{eqnarray}
\frac{dS}{dt} &=& \frac{d}{dt}\left\{-\frac{1}{p_{0}} \left( p_{0} \ln p_{0} 
+ (1 - p_{0}) \ln (1 - p_{0}) \right) \right\} \nonumber \\
&=& \ldots \, \, = \frac{\dot{p_{0}}}{p_{0}^{2}} \ln (1 - p_{0}),
\label{dsdt1}
\end{eqnarray}
 with the overdot denoting a derivative with respect to $t$. Combining 
eqns.(\ref{p01}) and (\ref{dsdt1}) we find, {\em for all} $t$,
\begin{equation}
\frac{dS}{dt} = \frac{\dot{\alpha}}{2 \alpha^{2}} \ln \left\{\frac{1 - \alpha}
{1 + \alpha} \right\}.
\end{equation}
Therefore, both the entropy, $S(t)$, and its rate of change, $dS/dt$, are 
determined entirely by the 
time-dependent coefficients $\Gamma_{1}$ and $\Gamma_{3}$ which arise in the
expression for the reduced density matrix, eqn.(\ref{rhor}). (Note that these 
are both independent of the centre of the initial minimum uncertainty
wavepacket, $(x_{0}, p_{0})$.) 

The asymptotic expression for $D$ given by eqn.(\ref{dasym}), 
along with that of eqn.(\ref{b2asym}), gives the asymptotic
behaviour of $\Gamma_{3}$ defined in eqn.(\ref{Gam3}):
\begin{equation}
\lim_{\kappa t \to \infty} \Gamma_{3} = \frac{8 \epsilon \kappa^{2}}{D_{asym}} 
\exp \left(-2(\kappa - \gamma_{0})t \right).
\label{Gam3asym}
\end{equation}

Inspection of the asymtotics of the various environmental terms used to 
define $\Gamma_{1}$ in eqn.(\ref{Gam1}), in particular eqns.(\ref{ta22asym}),
 (\ref{dasym}), (\ref{ta11asym}), (\ref{b3asym}), (\ref{b4asym}) and 
(\ref{ta12asym}), give   
\begin{eqnarray}
\lim_{\kappa t \to \infty} \Gamma_{1} &=& \frac{a_{22}(t\to\infty,0)}{\hbar} 
\nonumber \\
&=& \frac{\gamma_{0}}{2 \hbar} \coth \left(\frac{(\gamma_{0} + \kappa) \hbar}
{2 k_{B} T} \right),
\label{Gam1asym}
\end{eqnarray}
i.e., also a (temperature-dependent) {\em constant}. 

The asymptotic behaviour of $\alpha$ can now be determined using
eqns.(\ref{Gam3asym}) and (\ref{Gam1asym}):
\begin{equation}
\lim_{\kappa t\to\infty} \alpha = \xi \exp \left(-(\kappa - \gamma_{0})t \right),
\label{alphasym}
\end{equation}
where $\xi$ is a temperature-dependent, but time-independent constant.   
Clearly, $\alpha \to 0$ as $t \to \infty$, which means that $p_{0} \to 0$ 
too. Considering eqn.(\ref{dsdt1}) as $t \to \infty$ we find 
\begin{eqnarray}
\frac{dS}{dt} &=& \frac{\dot{p_{0}}}{p_{0}^{2}}\left(-p_{0} + p_{0}^{2}/2 +
\ldots \right) \nonumber \\
&\approx& -\frac{\dot{p_{0}}}{p_{0}} \nonumber \\
&=& \frac{-\dot{\alpha}}{\alpha(1 + \alpha)},
\end{eqnarray}
where we have used eqn.(\ref{p01}) in the last step. Finally, if we use the
asymptotic expression for $\alpha$, eqn.(\ref{alphasym}), we find
\begin{eqnarray}
\frac{dS}{dt} &\stackrel{\kappa t \to \infty}{=}& \kappa - \gamma_{0} 
\nonumber \\
&=& \kappa' \label{entrate}.
\end{eqnarray}
This result gives the asymptotic rate of entropy increase in 
situations where energy dissipation is important and cannot be neglected. As 
$\gamma_{0}$ determines the strength of the coupling to the environment we can 
now see that the rate at which an unstable, possibly chaotic system will lose 
information to the environment {\em does} depend on the coupling strength 
\cite{Zurek3}. 
Note, however, that this asymptotic rate does reduce to the asymptotic rate 
found previously in the weakly coupled regime
\begin{eqnarray}
\frac{dS}{dt} &\stackrel{\kappa t \to \infty}{=}& 
\sqrt{\lambda^{2} + \gamma_{0}^{2}} - \gamma_{0} \nonumber \\
&\approx& \lambda, \, \, \, \mbox{ when } \gamma_{0} \ll \lambda,
\end{eqnarray}
as required.

\section{Conclusion}

In this paper we have used the inverted harmonic oscillator to model
instability in open quantum systems. We have found that both the survival
probability function and the von Neumann entropy increase depend, for large
times, on the degree of instability in the system {\em and} on the strength 
of interaction with the environment in a simple way. The purpose of studying 
such 
an elementary system is to build up some degree of intuition as to the
behaviour of quantum chaotic systems coupled to an environment. A priori the 
claims of applicability of an inverted oscillator to modelling a chaotic 
system should be treated with caution. We will now discuss to what degree the
results for the oscillator can serve as a guide to actual quantum behaviour in
chaotic systems.

With regard to the survival probability function prediction we mention the 
study by D'Ariano et al. \cite{D'Ariano} of classical and quantum structures 
in the kicked-top model. No environment was included in their model but they 
have shown that the survival probability function (autocorrelation function in
their paper) provides an excellent means with which to compare classical and 
quantum invariant structures (periodic points), at least for predominantly 
regular kicked tops. This situation changes when most of the tori have
been destroyed. We are currently examining the influence a perturbed 
evolution has on this correspondence. Practically, the survival probability 
function can be obtained from an experimental spectrum 
(see Heller in \cite{Houches}, also \cite{spfexp}) and is a useful concept in 
studies of molecular
spectra. The longest events in the time domain determine the broadening of 
the peaks in the spectrum for an unstable periodic orbit, with the width being
given by the classical Lyapunov exponent. Our result, eqn.(\ref{spfasym}), 
suggests that this may change in the presence of an environment due to 
$\gamma_{0}$.      

For the entropy production prediction we mention two studies 
of the quantum kicked rotor model \cite{QC,Reichl,Casati} evolving as 
it interacts with a perturbing environment \cite{Miller,Zarum}. In the first 
\cite{Miller}, the constant von Neumann entropy production rate was 
calculated for various values of the nonlinearity parameter, $K$, of the 
system. The Lyapunov exponent, $\lambda$ can can be approximated by 
$\lambda \approx \log(K/2)$, for large $K$ and a {\em linear} dependence of the 
constant entropy production rate on the Lyapunov exponent was found, i.e.
\begin{equation}
\dot{S} = a \lambda + b,
\end{equation}
where $a$ and $b$ are constants determined by the environment. Moreover, this 
behaviour is seen in the mixed phase spaces resulting from low values of 
$K$ provided one considers only {\em local} Lyapunov exponents to be relevant.
In the second study \cite{Zarum} an initial state is perturbed unitarily as 
it evolves in 
time with the possible perturbations at each time step being taken from an 
ensemble. Averaging over the possible realisations requires that the state 
of the system be described by a density matrix as it is mixed. In this way the 
entropy can increase. It is shown that the eventually constant rate at which 
an initial coherent state produces entropy depends to a great degree upon 
where in phase space the centre of the wavepacket is situated: the more chaotic
the area of initial localisation - as quantified by a locally averaged 
Lyapunov exponent - the larger the constant entropy production rate.

We conclude, therefore, that there is indeed some value in using the toy 
model considered in this paper and others \cite{Zurek2} as a model of
instability in open quantum systems. The conjectures which follow from it
should be tested in more systems which are classically chaotic.

\section*{Acknowledgements}

Paul A. Miller would like to thank the King's College London Association 
(KCLA) for
a postgraduate studentship.

\section*{Appendix A}

In this appendix we will give the explicit definitions of the functions
$f_{i}, \, 1,\ldots,6$, $d_{1}(n)$ and $d_{2}(n)$ used to write the 
finite-temperature expressions for each $a_{ij}$ in eqns.(\ref{finta11}) - 
(\ref{finta22}). We will also derive their large-time limits, 
eqns.(\ref{ta11asym}) - (\ref{ta22asym}). 

Once again we will choose constant coupling constants in the ohmic spectral
density of eqn.(\ref{ohmic}) and we will also assume a large, but finite
$\omega_{c}$. This allows us to write 
\begin{equation}
I(\omega,s,s') \approx \frac{2 \gamma_{0} \omega}{\pi}.
\label{Iapprox}
\end{equation}
To calculate the finite temperature noise kernel of eqn.(\ref{nudef}) we will
now {\it formally} expand the $\coth$ function in a power series \cite{AS} 
and integrate term by term. Formally then
\begin{equation}
\nu(s,s') = \frac{2 \gamma_{0}}{\pi} \sum_{n=0}^{\infty} \frac{2^{2n} B_{2n}}
{(2n)!} \left(\frac{\hbar}{2k_{B}T}\right)^{2n-1} \int_{0}^{\infty} d\omega
\, \omega^{2n} \cos \omega(s-s').
\end{equation}
$B_{2n}$ are Bernoulli numbers. Defining
\begin{equation}
f(n) := \frac{2^{2n} B_{2n}}{(2n)!} \left(\frac{\hbar}{2k_{B}T}\right)^{2n-1}
\label{fn}
\end{equation}
we can rewrite this formal expansion as 
\begin{equation}
\nu(s,s') = 2 \gamma_{0} \sum_{n=0}^{\infty} f(n) \delta^{(2n)}(s-s'),
\end{equation}
i.e. as an infinite sum of derivatives of the $\delta$-function.
This expression can now be used in eqn.(\ref{acoeffs}), with eqns.(\ref{v1})
and (\ref{v2}), to calculate each $a_{ij}$. Writing $\hat{\lambda} :=
\lambda / \kappa$ we find
\begin{eqnarray}
a_{11}(z,0) &=& \frac{\hatgam}{(2 \hat{\lambda} \sinh z)^{2}}
\sum_{n=0}^{\infty} f(n) \left(\frac{d_{1}(n) f_{1}(z)}{\hatgam} + 
d_{2}(n) f_{2}(z) \right), \\     
a_{12}(z,0) &=& \frac{e^{-\hatgam z}}{2 (\hat{\lambda} \sinh z)^{2}}
\sum_{n=0}^{\infty} f(n) \left(d_{1}(n) f_{3}(z)- d_{2}(n) f_{4}(z) \right), 
\\   
a_{22}(z,0) &=& \frac{e^{-\hatgam z}}{(2 \hat{\lambda} \sinh z)^{2}}
\sum_{n=0}^{\infty} f(n) \left(d_{1}(n) f_{5}(z) + 
d_{2}(n) f_{6}(z) \right), 
\end{eqnarray}
where
\begin{equation}
d_{\left\{{1 \atop 2}\right\}}(n) := \frac{\kappa^{2n}}{2} 
\left((\hatgam - 1)^{2n} \pm (\hatgam + 1)^{2n} \right),
\end{equation}
and
\begin{eqnarray}
f_{1}(z) &:=& \hatgam^{2} \cosh 2z + \hatgam \sinh 2z - \exp(2\hatgam z) + 1 -
\hatgam^{2}, \\
f_{2}(z) &:=& \hatgam \sinh 2z + \cosh 2z - \exp(2\hatgam z), \\
f_{3}(z) &:=& \cosh z (\exp(2\hatgam z) - 1) - \hatgam \sinh z 
(\exp(2\hatgam z) + 1), \\                             
f_{4}(z) &:=& \hatgam \cosh z (1 - \exp(2\hatgam z)) + \sinh z 
(\exp(2\hatgam z) - 1) \nonumber \\ & & + 2 \hatgam^{2} \sinh z, \\ 
f_{5}(z) &:=& 1 - \hat{\lambda}^{2} \exp(2\hatgam z) - 
\hatgam \exp(2\hatgam z) \left(\hatgam \cosh 2z - \sinh 2z \right), \\
f_{6}(z) &:=& \hatgam \left( \exp(2\hatgam z) (\hatgam \sinh 2z - \cosh 2z) 
+ 1 \right). 
\end{eqnarray}

\subsection*{Asymptotics of $a_{ij}$}

It is easy to show that 
\begin{equation}
\frac{f_{1}(z)}{\sinh^{2} z} \stackrel{z \to \infty}{\longrightarrow}
2 \left( \hatgam^{2} + \hatgam \right)
\end{equation}
and
\begin{equation}
\frac{f_{2}(z)}{\sinh^{2} z} \stackrel{z \to \infty}{\longrightarrow}
2 \left( \hatgam + 1 \right).
\end{equation}

These expressions allow us to determine the asymptotic behaviour of
$a_{11}(z,0)$ defined in eqn.(\ref{finta11}). We find
\begin{eqnarray}
\lim_{z \to \infty} a_{11}(z,0) &=& \frac{\hatgam}{4 \hat{\lambda}^{2}}
\sum_{n=0}^{\infty} f(n) \left( \frac{d_{1}(n)}{\hatgam}\left[2(\hatgam^{2} +
\hatgam) \right] + d_{2}(n) \left[2 (\hatgam + 1) \right] \right) \nonumber \\
&=& \frac{\hatgam(\hatgam + 1)}{2 \hat{\lambda}^{2} }
\sum_{n=0}^{\infty} f(n) \left(\hatgam - 1 \right)^{2n} \kappa^{2n}.
\end{eqnarray}
One next uses the definition of $f(n)$ in eqn.(\ref{fn}) and the formal power
series expansion of the $\coth$ function derive the desired 
expression given in eqn.(\ref{ta11asym}) below. 
To derive eqns.(\ref{ta12asym}) and (\ref{ta22asym}) we proceed along similar 
lines:       
\begin{eqnarray}
a_{11}(z,0) & \to & - \frac{\hatgam \kappa}{2} \coth \left(\frac{(\gamma_{0} - 
\kappa) \hbar}{2 k_{B} T} \right), \label{ta11asym}  \\        
a_{12}(z,0) & \to & \kappa e^{-(1-\hatgam)z} \coth \left(\frac{(\gamma_{0} + 
\kappa) \hbar}{2 k_{B} T} \right), \label{ta12asym} \\        
a_{22}(z,0) & \to & \frac{\hatgam \kappa}{2} \coth \left(\frac{(\gamma_{0} + 
\kappa) \hbar}{2 k_{B} T} \right). \label{ta22asym}
\end{eqnarray}

\end{document}